\documentclass[twocolumn,english,aps,prb,showpacs,byrevtex,amsmath,amssymb,superscriptaddress]{revtex4-1}
\usepackage[T1]{fontenc}
\usepackage[latin9]{inputenc}
\usepackage{textcomp}
\usepackage{amsmath}
\usepackage{graphicx}

\makeatletter
\usepackage{lipsum}
\usepackage{epsfig}
\usepackage{subfigure}
\usepackage{color}

\definecolor{Red}{rgb}{1,0,0}
\definecolor{Blu}{rgb}{0,0,1}
\definecolor{Green}{rgb}{0,1,0}

\makeatother

\usepackage{babel}
\begin{document}

\title{First principles study of structural, magnetic and electronic properties of CrAs}

\author{Carmine Autieri}

\affiliation{CNR-SPIN, UOS L'Aquila, Sede Temporanea di Chieti, I-66100 Chieti, Italy}

\author{Canio Noce}

\affiliation{CNR-SPIN, I-84084 Fisciano (Salerno), Italy}

\affiliation{Dipartimento di Fisica ``E. R. Caianiello'', Universit\`a di
	Salerno, I-84084 Fisciano (Salerno), Italy}

\date{\today}
\begin{abstract}
We report {\it ab initio} density functional calculations of the structural and magnetic properties, and the electronic structure of CrAs. To simulate the observed pressure-driven experimental results, we perform our analysis for different volumes of the unit cell, showing that the structural, magnetic and electronic properties strongly depend on the size of the cell. We find that the calculated quantities are in good agreement with the experimental data, and we review our results in terms of the observed superconductivity.
\end{abstract}


\maketitle

\section{Introduction}
\begin{figure}
	\centering
	\includegraphics[width=8.6cm,height=6.3cm,angle=0]{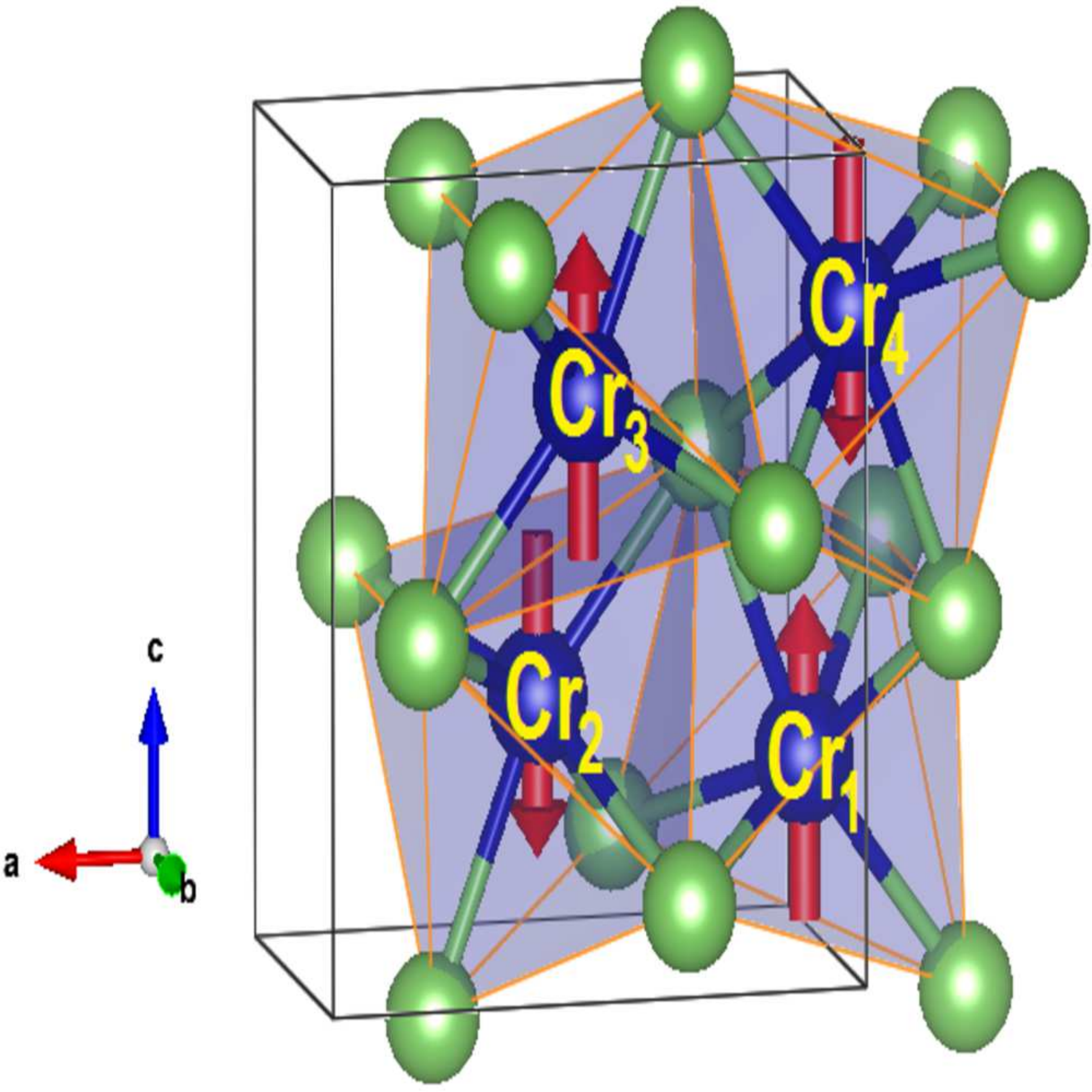}
	\caption{Crystal structure of the CrAs. Cr and As are shown as blue and green spheres, respectively. All the Cr atoms, labelled as 1, 2, 3, 4, are equivalent even though the Cr-Cr distances are different. The red arrows represent the spin of the Cr atoms in the magnetic ground state for the collinear approximation.
	}
	\label{Magnetism}
\end{figure}

Over the last decades, the investigation of the interplay of superconductivity and magnetism has received great attention.~\cite{norman11} It has been realized, indeed, that the proximity to a magnetic phase could lead to new forms of superconductivity with non-conventional order parameter.~\cite{goll06}  The systems where these effects have been detected include the heavy-fermion materials,~\cite{movshovich01} the organic superconductors,~\cite{lefebvre00} the cuprate ~\cite{vanharlingen95} and the ruthenate superconductors~\cite{mackenzie03} as well as the iron-based materials.~\cite{mazin08}
As a general trend, in these materials the superconductivity emerges near a magnetic quantum critical point (QCP) where a high-temperature ordered state involving spin, charge or lattice degrees of freedom is suppressed by applying external tuning parameters, such as the charge doping, the chemical substitutions or the external pressure.~\cite{norman11} Moreover, the temperature T$_c$ where the transition to the superconducting phase takes place, always passes through a maximum value, at some critical value of the control parameter, leading to well-known dome-shaped T$_C$ versus tuning parameter phase diagram.~\cite{varma99,noce00,noce02,vandermarel03,jiang09,shibauchi14,seo15} This property has suggested that seeking for a QCP should provide an effective approach for searching new classes of unconventional superconductors.
This behaviour is well illustrated by the recent discovery of pressure-induced superconductivity in CrAs, the first Cr-based unconventional superconductor.~\cite{wu14}  CrAs belongs to the large family of transition-metal pnictides with a general formula MX (M=transition metal, X=P, As, Sb), which form in either hexagonal NiAs-type B8$_1$ phase (space group P6$_3$/mmc) or orthorhombic MnP-type B31 phase (space group Pnma) structure.~\cite{wilson64} Indeed, the compound exhibits a phase transition at 800 K from the hexagonal NiAs-type to the orthorhombic MnP-type configuration. In this latter phase, shown in Fig.~\ref{Magnetism},  the unit-cell lattice parameters are $a$=5.649 {\AA}, $b$=3.463 {\AA} and $c$=6.2084 {\AA}.~\cite{wu14} We notice that the Cr atoms are located in the centre of CrAs$_6$ octahedra, surrounded by six nearest-neighbour arsenic atoms, and four of the six Cr-As bonds are inequivalent due to the high anisotropy exhibited by this compound.

Similar to parent iron pnictide superconductors, the neutron scattering experiments reveal that below 270 K, CrAs is magnetically ordered with a double helix propagating along the $b$-axis. The helimagnetic phase is suppressed at about 0.7 GPa, and the superconductivity is realized in a wide pressure range, in the non magnetic state, with a maximum T$_c\sim 2.2\,$K at 1.0 GPa.~\cite{kotegawa14,kotegawa15,khasanov15,keller15}
The structure of CrAs remains stable up to 1.8\, GPa, whereas the lattice parameters exhibit anomalous compression behaviors.  By increasing the pressure, the lattice parameters $a$ and $c$ both show a non-monotonic change whereas the lattice parameter $b$ undergoes a rapid contraction at $\sim$0.18-0.35~GPa, which suggests a pressure-induced isostructural phase transition.~\cite{wu14,keller15}
The T$_c$ vs P phase diagram of CrAs resembles that of many of the unconventional superconducting systems above mentioned suggesting the presence of strong antiferromagnetic fluctuations near the critical pressure P$_c$ and a possible unconventional pairing mechanism.~\cite{wu14}

For the sake of completeness we mention that detailed studies of the stability of the ferromagnetism in CrAs crystallizing in the zinc-blende structure have been reported.~\cite{sasoglu05} The electronic structure is calculated from first principles, whereas the frozen-magnon approximation is employed to evaluate the exchange
parameters and the Curie temperature. It is shown that the CrAs exhibits a very high
Curie temperature, and the ferromagnetism is stable with respect to the compression of the lattice parameters.

The purpose of this paper is twofold: (1) we will present the structural and magnetic properties as well as the electronic structure of this superconducting material, here investigated by means of {\it ab initio} calculations using density functional theory, and (2) we will discuss the effect of the external pressure on these quantities. The paper is organized as follows: in the next section we will provide some details on the computational approach adopted; Sec. III will be devoted to the investigation of the structural  properties of the compound; in Sec. IV we will discuss the magnetic properties,  while in Sec. V we will present the results for the electronic structure, with special emphasis addressed to the density of states (DOS) and Fermi surface (FS). Sec. VI is devoted to the discussion of the results and, finally, the last Section contains a summary together with the conclusions and some remarks and speculations about the interplay between the results here presented and the superconductivity exhibited by the CrAs.

\section{Computational details}
We have performed first-principles density functional theory (DFT) calculations by using the VASP package, version 5.4.1.~\cite{VASP2}
The core and the valence electrons were treated within the Projector Augmented Wave (PAW) method~\cite{VASP} with a cutoff of 400~eV for the plane wave basis. After standard tests, we have used an accurate PAW with 14 valence electrons for the Cr (3s$^2$3p$^6$4s$^2$3d$^4$) and 5 valence electrons for the As (4s$^2$4p$^3$).
All the calculations have been performed using a 12$\times$16$\times$10 $k$-point Monkhorst-Pack grid.~\cite{Monk} For the treatment of exchange-correlation, the Local Spin Density Approximation (LSDA) and the Perdew-Zunger~\cite{Perdew} parametrization of the Ceperly-Alder~\cite{Ceperley} data have been implemented.

We have tested other {\it ab initio} approaches such as
the spin generalized gradient approximation (SGGA),\cite{PBE} the SGGA+U and the LSDA+U schemes. These methods produce a magnetization larger and more far from the experimental results than that obtained by means of LSDA approach, although the SGGA method gives better results for the volume estimation.
We notice that a similar problem has been already emphasized investigating the magnetic properties of metallic Sr-ruthenates.\cite{Autieri2016}
Therefore, we decide to use the LSDA method since it gives a reasonable agreement with the magnetic experimental data (see Section IV) even though the structural properties are not described in an excellent way.

In all cases the tetrahedron method with Bl\"{o}chl corrections~\cite{BlochlCORR} has been used for the Brillouin zone integrations. We have optimized the internal degrees of freedom by minimizing the total energy to be less than 7$\times10^{-7}$ eV. Such precision is necessary due to the presence of several different magnetic phases close to the ground state. After obtaining the Bloch wave functions $\psi_{n,\textbf{k}}$, the Wannier functions~\cite{Marzari97,Souza01} are build up using the WANNIER90 code~\cite{Mostofi08} generalizing the following formula to get the Wannier functions $W_n(\textbf{r})$:
\begin{equation}
  W_n(\textbf{r})=\frac{V}{(2\pi)^3}\int d\textbf{k} \psi_{n,\textbf{k}}e^{-i\textbf{k}\cdot\textbf{r}}\, ,
\end{equation}
were $V$ is the volume of the unit cell and $n$ is the band index.

To extract the low energy properties of the electronic bands, we have used the Slater-Koster interpolation scheme. In particular, we have fitted the electronic bands, and in this way we have been able to get the hopping parameters and the spin-orbit constants. This approach has been applied to determine the real space Hamiltonian matrix elements in the maximally localized Wannier function basis and to find out the Fermi surface.

For completeness, we mention that a way to possibly improve the results of this paper could be the application of some hybrid functionals. Nevertheless, while some hybrid exchange functionals are not suitable to describe the metals,~\cite{Kresse2007} the B3PW91 code has been demonstrated to give good results to describe the magnetic metals,~\cite{Yaz} giving rise to a magnetization lower than that obtained by means of SGGA when low values of the mixing parameters are considered.

\section{Structural properties}
To investigate the structural properties such as the lattice constants, the bulk modulus, and the total energy, the numerical data were fit to the Murnaghan-Birch equation of state.~\cite{birch47}	

Since LSDA strongly underestimates the equilibrium volume whereas the Perdew, Burke, Ernzerhof (PBE) gradient approximation approach~\cite{PBE} slightly overestimates this quantity, to evaluate the structural properties we will use the PBE formulated for solids (PBEsol).~\cite{PBEsol}
To determine the structural properties in the ground state, as previously said, we use the Birch-Murnaghan equation of state:~\cite{birch47}
\scriptsize
\[
E(V)=E_0 +
\]
\begin{equation}\label{eq:BIRMUR2}
\frac{9V_0B_0}{16}\left\lbrace \left[\left( \frac{V_0}{V}\right)^{\frac{2}{3}}-1\right]^3 B_0^\prime + \left[\left( \frac{V_0}{V}\right)^{\frac{2}{3}}-1\right]^2 \left[6-4\left(\frac{V_0}{V}\right)^{\frac{2}{3}}\right]\right\rbrace \, ,
\end{equation}
\normalsize
where $V_0$ and $E_0$ are the equilibrium volume and the total energy at the equilibrium volume, respectively; $B_0$ is the bulk modulus, and $B_0^\prime$ its pressure derivative.	
	
These parameters are usually obtained from fits to experimental data. Thus, using the DFT data and Eq.~(\ref{eq:BIRMUR2}), in the Table~\ref{TabSTRUCT} we report the results of this fitting procedure together with the experimental lattice constants and the corresponding volume, for two values of the external pressure, namely the ambient pressure and  P=0.6~GPa. We find that the equilibrium volume is 117.0~\AA$^3$, which is comparable with the corresponding experimental values. Moreover, the bulk modulus and its derivative we found are 119~GPa and 4.5, respectively, whereas the formation enthalpy is -826~KJ/mol.
\begin{table}[!ht]
	\caption{Structural properties within the PBEsol scheme compared with the experimental data at different pressures. The equilibrium volume $V_0$ is expressed in \AA$^3$ unit. The lattice constants $a$, $b$, and $c$ are expressed in \AA. The pressure is given in GPa unit.}
	\begin{center}
		\renewcommand{\arraystretch}{1.7}
		\begin{tabular}{|c|c|c|c|}
			\hline
			               &   PBEsol  &  P=0 &  P=0.6 \\
			\hline
			V$_0$          &    117.0  &  123.4   &   121.7 \\
			\hline
			a              &    5.498  &  5.649   &   5.570 \\
			\hline
			b              &    3.524  &  3.463    &  3.570 \\
			\hline
			c              &    6.039  &  6.208    &  6.118  \\
			\hline
			
		\end{tabular}
	\end{center}
	\label{TabSTRUCT}
\end{table}

\begin{figure}
	\centering
	\includegraphics[width=8.6cm,height=6.3cm, angle=0]{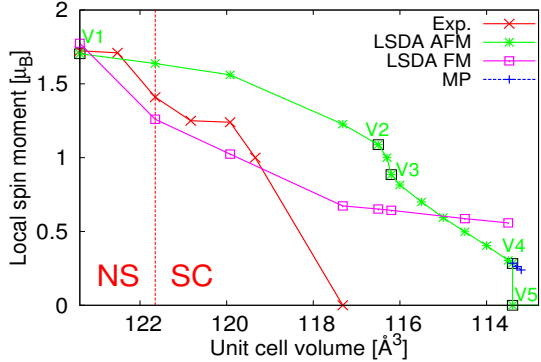}
	\caption{Evolution of the local magnetic moment of the Cr atoms as a function of the volume of the unit cell
		within the LSDA  for the AFM phase (green line), FM phase (pink line) and experimental results~\cite{expvolume} (red line).
		A metastable phase (MP), represented as blue points, has been found at high pressure. We point out that at zero pressure the system is in the normal state (NS). The vertical dashed red line indicates the volume where the superconducting state (SC) comes in. The values of volumes V1-V5 are reported in the text.}
	\label{Moment}
\end{figure}

\section{Magnetic properties}
To investigate the magnetic properties of CrAs, when the external pressure is varied, we have used the atomic positions and the volumes reported in Ref.~\onlinecite{expvolume}, assuming the orthorhombic MnP-type structure since it is the phase adopted by the CrAs at low temperatures. To scrutinise the magnetization at lower volumes we have performed a further uniform compression of the lattice constants. In this way we have performed the calculations for the volumes V1-V5, reported in Fig.~\ref{Moment}, where V1=123.4  {\AA}$^3$, V2=116.5  {\AA}$^3$, V3=116.2  {\AA}$^3$ and V4=V5=113.5  {\AA}$^3$. We point out that the smallest volume of the unit cell for which the experimental data are available is 117.32 {\AA}$^3$.

The variation of the magnetic moment as a function of the volume is reported in Fig.~\ref{Moment}, where we plot the experimental available results~\cite{expvolume} together with our theoretical predictions. In particular, we have calculated the magnetic moment applying the LSDA for both the antiferromagnetic case and the ferromagnetic one. Looking at this figure, we infer that the calculations suggest an antiferromagnetic ground state. Indeed, the trend of the magnetic moment, in the ferromagnetic configuration, does not follow the experimental results even though the numerical values of the magnetic moments are comparable to the experimental ones.
Concerning the antiferromagnetic results, we can observe a discernible change of the magnetic moment for the volume V5. This variation corresponds to a first order phase transition between two magnetic phases having different magnetic moment. A metastable phase is also found at the same pressure. When V is equal to V3 the magnetic moment exhibits another discernible change. Having performed an accurate calculation with a large number of $k$-points, we can state that this change is not a first order phase transition but an abrupt variation of the magnetic moment. For completeness, we mention that the presence of several different magnetic phases close to the ground state was already found in other structural phases of the CrAs.~\cite{PHASETRANSITION1,PHASETRANSITION2}
As a final consideration, we would like to point out that the reasonable values found for the magnetic moment without the inclusion of the Coulomb repulsion suggest that the CrAs may be considered as a weakly correlated metallic itinerant antiferromagnet, with a magnetic moment very sensitive to the chosen volume and to the magnetic configuration adopted.

Although the CrAs cannot be considered as a Heisenberg antiferromagnet, we estimate the magnetic exchange interaction to calculate the order of magnitude of the magnetic couplings.
Moreover, we find out that there are two very different Cr-Cr distances along the $c$-axis, namely 3.090 and 4.042 {\AA}, corresponding to Cr$_2$-Cr$_3$ and Cr$_1$-Cr$_4$ distances in Fig.~\ref{Magnetism}, respectively.
Then, we have calculated the magnetic couplings considering the energy differences between different magnetic configurations.
The magnetic coupling between the Cr atoms with shorter distance is strongly antiferromagnetic (-60 meV) while the coupling between Cr atoms with longer distance is weakly ferromagnetic (+10 meV).
On the other hand, the magnetic coupling along the $a$-axis is weakly antiferromagnetic (-10 meV). Hence, the competition between these magnetic couplings may be responsible for a magnetic frustration giving rise to the complex magnetic structure experimentally detected. We also point out that in Refs.~\onlinecite{expvolume,Magnetism} it is reported that the magnetic moment of the Cr atoms with shorter distance are antiparallel, indicating a strong antiferromagnetic coupling between these Cr atoms, in agreement with our results.
Moreover, the magnetic ground state we find out is a $G$-type antiferromagnetic state whose behavior is close to the ground state experimentally reported.

\begin{figure}[t!]
	\centering
	\includegraphics[width=8.6cm,height=6.3cm, angle=0]{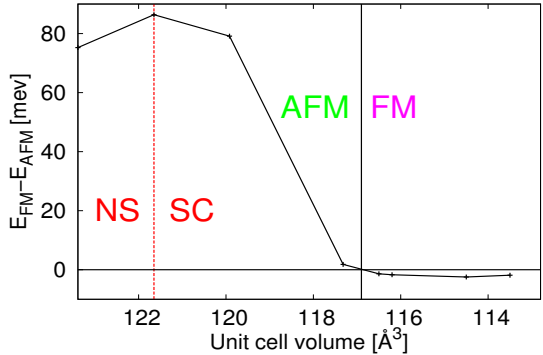}
	\caption{Energy difference E$_{FM}$-E$_{AFM}$ per formula unit as function of the volume. In the superconductive region the energy difference gets reduced and the system becomes ferromagnetic.}
	\label{DeltaE}
\end{figure}

Furthermore, we have evaluated the difference between the energy of the ferromagnetic phase $E_{FM}$ and that of the $G$-type antiferromagnetic phase $E_{AFM}$, this latter quantity being related to the N\'{e}el critical temperature $T_N$. In the Fig. \ref{DeltaE} we show the evolution of this energy difference as a function of the volume. We find that $E_{FM}$-$E_{AFM}$ is +75.25~meV, per formula unit, at zero pressure while its minimum value is -2.46~meV.	

For volumes larger than 118 \AA$^3$, the system is an antiferromagnet with large T$_N$, whereas in the experimental superconducting region below 118 \AA$^3$ the magnetic exchange couplings are reduced by a factor 40 respect to the zero pressure case. Below 117 \AA$^3$ the system becomes ferromagnetic with very small exchange coupling.
As plotted in Fig.2, a non vanishing magnetic moment is observed at very low volumes but the magnetic order is very week. We also point out that the ferromagnetic ground state configuration appears at volumes below the smallest available experimental volume.
Interestingly, we do not have strong evidence of a ferromagnetic phase that could bring towards a triplet superconductivity in a region of the phase diagram.

\section{Electronic properties}

\subsection{Band structure}

\begin{figure}[b!]
	\centering
	\includegraphics[width=8.6cm,height=6.3cm, angle=0]{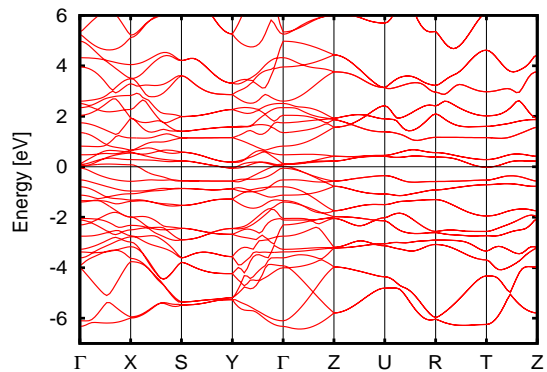}
	\caption{Band structure of CrAs at the volume V1 in the energy range [-7.0, +6.0] eV
		along the high-symmetry path $\Gamma$-X-S-Y-$\Gamma$-Z-U-R-T-Z.~\cite{setyawan10} The horizontal
		line denotes the Fermi energy.
	}
	\label{BS1}
\end{figure}

\begin{figure}[t!]
	\centering
	\includegraphics[width=8.6cm,height=6.3cm, angle=0]{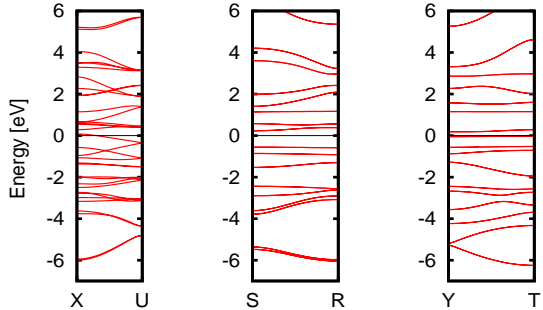}
	\caption{The same of Fig.~\ref{BS1} along three different vertical high-symmetry paths:
		X-U, S-R and Y-T.~\cite{setyawan10}
    }
	\label{BS2}
\end{figure}

To discuss the band structure of CrAs, in this Section we describe and comment on, as a representative example, the band structure only for the largest volume V1 (see Figs.~\ref{BS1} and \ref{BS2}). The high symmetry points along which we plot the band structure have been chosen according to the notation quoted in Ref.~\onlinecite{setyawan10}.

First of all, we point out that we have found that this compound does not show any relevant charge transfer, and the oxidation state is basically zero both for Cr and As ions. The Cr atoms are in a d$^6$ configuration while the As atoms are in a p$^3$ configuration.
Moreover, there are flat bands between -2.0 and +2.0 eV and wider bands out of this range.
The presence of flat bands around the Fermi level gives rise to van Hove singularities, as may be easily inferred also from an inspection of the DOS for the volume here considered (see next Section).
Furthermore, we can also infer a strong anisotropy as observed from the band structure along the $\Gamma$-X, $\Gamma$-Y and $\Gamma$-Z paths.
The dominant hopping parameters are the first neighbor Cr-As hoppings, while the Cr-Cr hopping parameters along y and z directions are smaller than the corresponding ones along the x direction. However, along the x direction the Cr atoms are shifted so that we find relevant non-diagonal hopping in the cubic harmonic basis, while the nondiagonal hopping in the y and z directions are negligible.

In the non-magnetic (NM) and antiferromagnetic (AFM) phases, the spin up channel band structure is equal to the spin down channel band structure due to the inversion and time reversal symmetries. The band structure of the orthorhombic MnP-type phase exhibits an additional degeneracy at the edge of the first Brillouin zone in both spin channels.~\cite{BS_MnP} However, in the AFM phase we find that the degeneracy is broken along the XS line. Along the XS direction, the bands are produced by a mixing of Cr$_1$ with Cr$_3$ sites and a mixing of Cr$_2$ with Cr$_4$ sites. This magnetic configuration gives rise to bands with minority and majority electrons, whereas the removal of the degeneracy is a consequence of the spin-splitting between these bands. Therefore, this region of the energy spectrum is strongly dependent on the magnetic moment.

\begin{figure}[t!]
\centering
\includegraphics[width=8.6cm,height=6.3cm, angle=0]{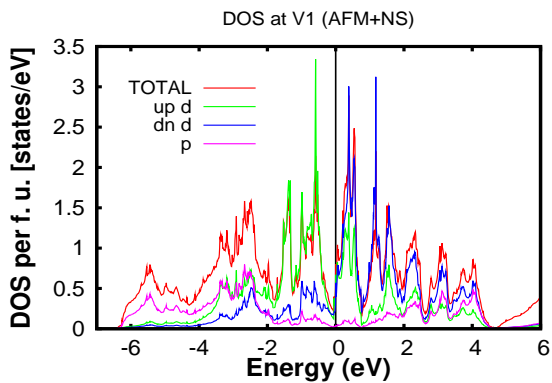}
\caption{DOS for the unit cell volume V1. The total DOS per formula unit is plotted as red.
The majority Cr states, minority Cr states and As-p states are plotted as green, blue and pink lines, respectively.
For this volume, the system is a non-superconducting antiferromagnet metal.}
\label{DOSV1}
\end{figure}

\begin{figure}[b!]
\centering
\includegraphics[width=8.6cm,height=6.3cm, angle=0]{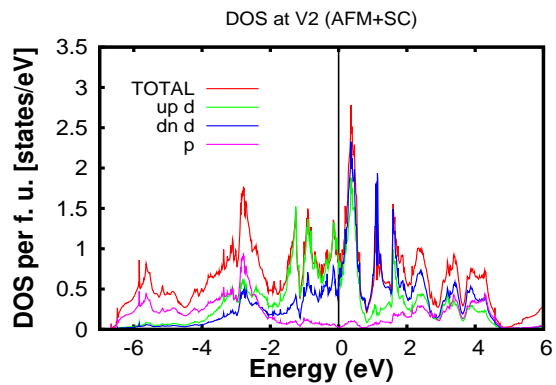}
\caption{DOS for the unit cell volume V2. The total DOS per formula unit is plotted as red.
The majority Cr states, minority Cr states and As-p states are plotted as green, blue and pink lines, respectively.
For this volume, the system is a superconducting antiferromagnet metal.}
\label{DOSV2}
\end{figure}

We have also calculated the spin-orbit coupling constant $\lambda$ for the Cr and As atoms  from the real space Hamiltonian. We have found that $\lambda_{Cr}$=33~meV for the $3d$ orbitals of the Cr atoms and $\lambda_{As}$=164~meV for the $4p$ orbitals of the As atoms. We notice that the spin-orbit coupling of Cr is in good agrement with the value (34~meV) found in an other anisotropic environment with monoclinic space group.~\cite{Autieri2014}
The presence of spin spirals in the CrAs magnetic structure suggests the presence of a large Dzyaloshinskii-Moriya  interaction between two magnetic atoms via a non magnetic atom.
The key ingredients for this interaction are the spin-orbit coupling and the Cr-As-Cr hybridization.
From the calculated values of the spin-orbit couplings, it is not clear whether the Dzyaloshinskii-Moriya interaction is mainly due to the SOC of the non magnetic atom\cite{Fert1980}, to the SOC of the magnetic atom\cite{Shanavas2016} or both.

\subsection{Density of states}

\begin{figure}[t!]
\centering
\includegraphics[width=8.6cm,height=6.3cm, angle=0]{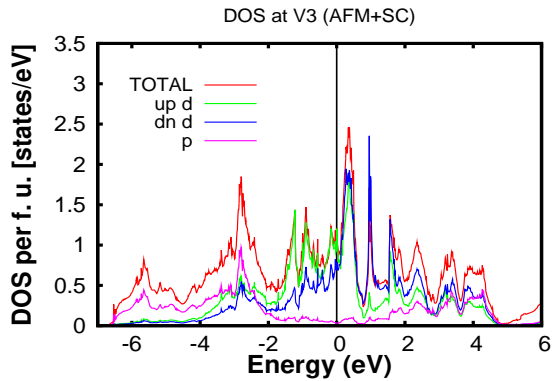}
\caption{DOS for the unit cell volume V3. The total DOS per formula unit is plotted as red.
The majority Cr states, minority Cr states and As-p states are represented by means of green, blue and pink lines, respectively. For this volume, the system is a superconducting antiferromagnet metal.
}
\label{DOSV3}
\end{figure}

\begin{figure}[b!]
\centering
\includegraphics[width=8.6cm,height=6.3cm, angle=0]{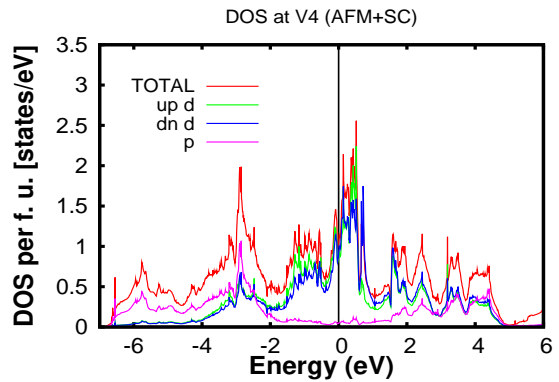}
\caption{ DOS for the unit cell volume V4. The total DOS per formula unit is plotted as red.
The majority Cr states, minority Cr states and As-p states are represented by means of green, blue and pink lines, respectively. For this volume, the system is a superconducting antiferromagnet metal.
}
\label{DOSV4}
\end{figure}

\begin{figure}[b!]
\centering
\includegraphics[width=8.6cm,height=6.3cm, angle=0]{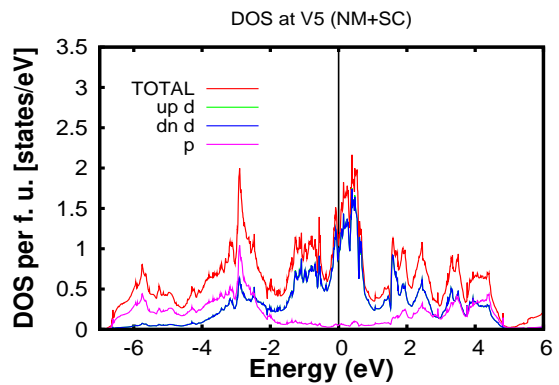}
\caption{ DOS for the unit cell volume V5. The total DOS per formula unit is plotted as red.
The majority Cr states, minority Cr states and As-p states are represented by means of green, blue and pink lines, respectively.
For this volume, the system is a superconducting non-magnetic metal.
}
\label{DOSV5}
\end{figure}

Since the CrAs exhibits interesting magnetic features for unit cell volumes equal to V1, V2, V3, V4 and V5 (see Fig.~\ref{Moment}), here we present the calculation of the DOS for these values of V. In this way we will investigate what happens close the magnetic transition V5-V4, near the jump V3-V2 and in the antiferromagnetic V1 phase. We notice that for V=V1 the CrAs in experimentally found in the normal state phase, whereas for the other volumes it is superconducting.\cite{wu14}  These DOSs are reported in Figs.~\ref{DOSV1}-\ref{DOSV5} with the Fermi level at 0 eV.

As a general trend, the DOS, from -6.5~eV to -3.5~eV, has a predominant As character, since in this energy range the bands are almost totally occupied by As states. On the other hand, from -3.5~eV to -2.0~eV we find a coexistence of Cr-d and As-p states, whereas from -2.0~eV to +2.0~eV we have essentially Cr bands that are more flat compared to the As bands. The coexistence of unoccupied Cr-d and As-p states is found in the range  [2.0,\, 4.5]~eV, and finally the bands that originate from 4s electrons of the Cr atoms are located above 4.5~eV.

Differently from transition metal oxides, for CrAs we cannot decouple the Cr bands from the As bands. In the Wannier analysis, we cannot disentangle the p-bands from the d-bands since they are not well separated in energy, and this property supports what we can infer from the DOS and the band structure. At the Fermi level, the character of the bands is predominantly due to Cr atoms, whereas the As weight is relevant below -2.0~eV and above +2.0~eV.
Moreover, we notice that the bands become more flat close to the Fermi level because the contribution from p-orbitals coming from As ion is quite small due to the fact that the As bands are mainly located few eV above and below the Fermi level.
Furthermore, we observe that the As states are more delocalized compared to the Cr states, and when the volume decreases the d-electron bands are more delocalized giving rise to an increase of the bandwidth.
We also remark the presence of the large peak close to the Fermi level, analogous to the FeSe and FeS superconductors.~\cite{DSingh}
From these figures it is easily inferred that the bandwidth is 11.0~eV large at the volume V1 while it is approximately  12.0~eV at the volume V5. We point out that when the volume of the system decreases from V1 to the volume V5, the compound becomes non-magnetic, and in the nonmagnetic V1 phase the bandwidth of the system is 11.3~eV large, indicating that the transition from antiferromagnetic to nonmagnetic phase is accompanied by an increase of the bandwidth of about 0.3~eV.

We would like to stress that for all the volumes investigated, we haven't observed a pseudogap phase and, although in other Cr/As superconducting materials it has been suggested a d$^4$ Cr configuration,~\cite{Pizarro} our calculations support a configuration with 5.5 Cr electrons similarly to hole doped pnictides.

\begin{figure}[t!]
\centering
\includegraphics[width=8.6cm,height=6.3cm, angle=0]{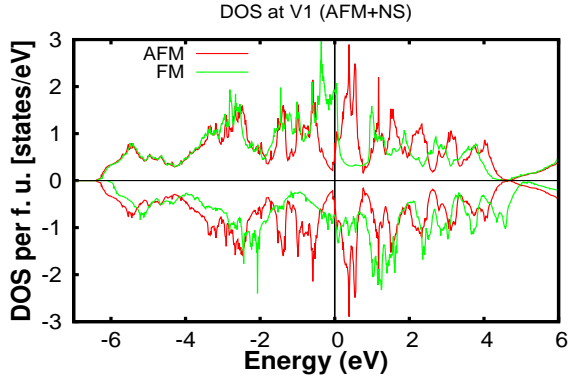}
\caption{DOS comparison for the unit cell volume V1.
 The total DOS per formula unit for the AFM (FM) is plotted as red (green).
The majority states (minority states) are plotted in the upper panel (bottom panel).}
\label{DOSFMV1}
\end{figure}

Finally, for completeness, in Fig.~\ref{DOSFMV1} we have compared the DOS for the AFM and FM phase at the volume V1. We see that while below -2~eV and above 1~eV the DOS are slightly affected by the magnetic ground state considered, due to the fact that these regions of the energy spectrum are mainly related to As states, the region between the previous values is strongly modified when the two magnetic configurations are considered. In the FM phase at the Fermi level the DOS for the majority spin channel is larger than the DOS of the minority spin channel and the bandwidth of the majority spin channel is roughly 0.2~eV smaller than the corresponding bandwidth of the AFM phase. On the contrary, the opposite situation happens for the minority spin channel bandwidth being the FM one larger than the AFM one of about 0.2~eV.

\subsection{Fermi surface}

\begin{figure}[t!]
\centering
\includegraphics[width=9cm,height=4.7cm,angle=0]{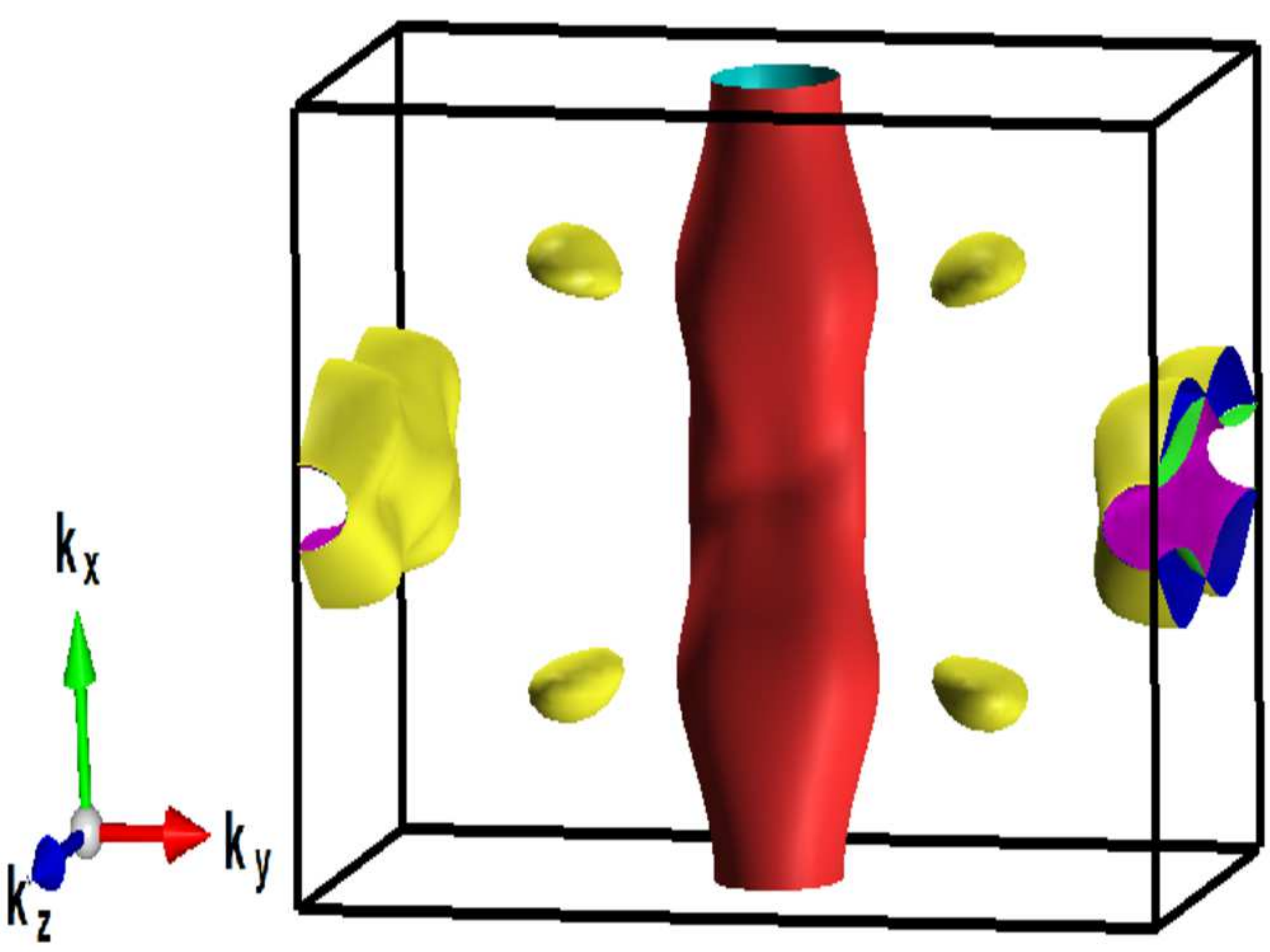}
\caption{Fermi surface of CrAs at the volume V1 in the first Brillouin zone.
}
\label{V1_FS}
\end{figure}

Since electrons close to the Fermi level are primarily responsible for superconductivity, the FS is a key quantity to investigate to understand the electronic structure of any metallic material. We point out that the FS can be experimentally probed, for example, using de Haas-van Alphen experiments.
Let us here study how the FS varies when the volume of the unit cell is modified, starting from the largest volume V1 (Fig.~\ref{V1_FS}) to the smallest one V5 (Fig.~\ref{V5_FS}).

\begin{figure}[b!]
\centering
\includegraphics[width=9cm,height=4.4cm,angle=0]{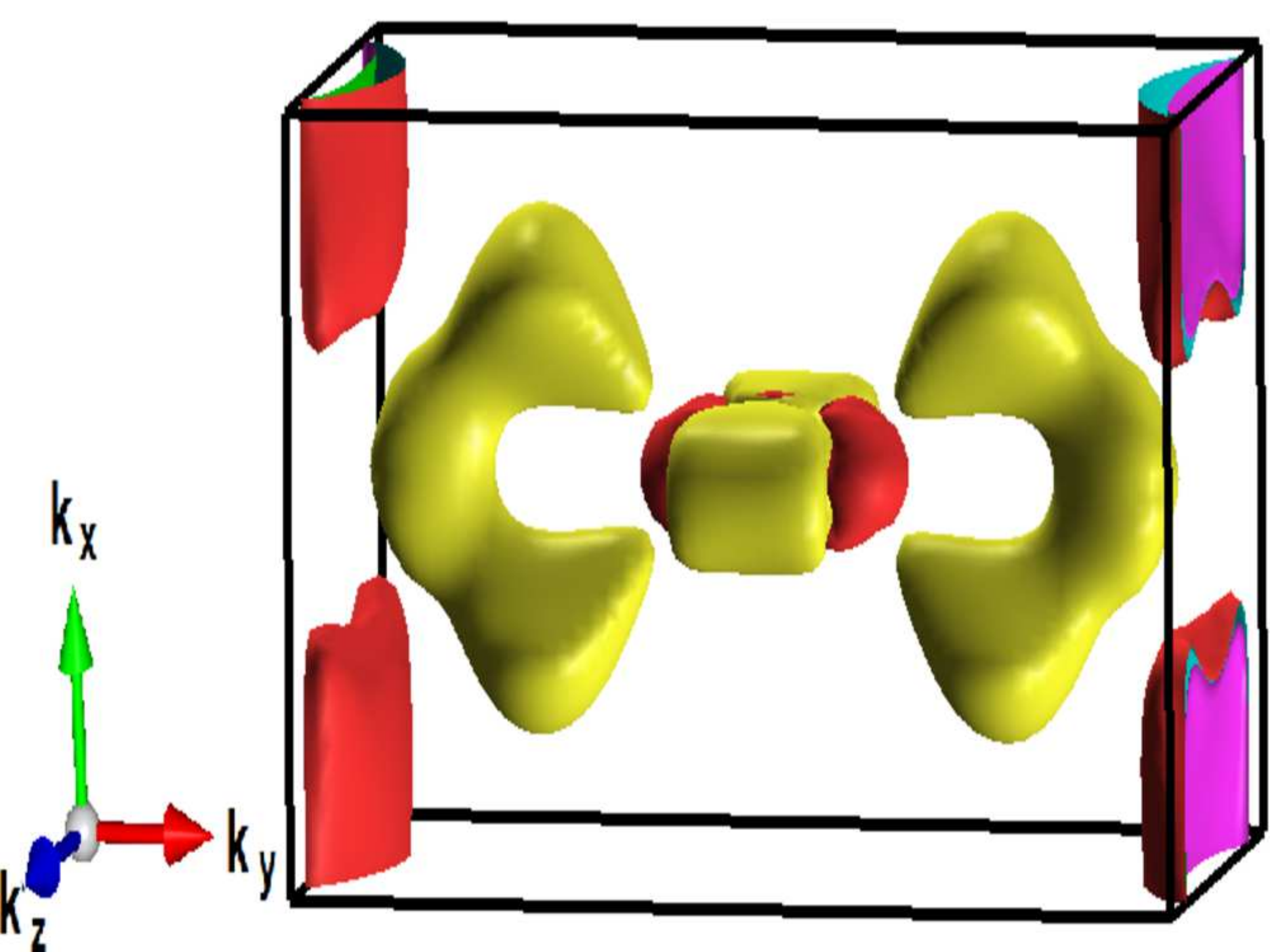}
\includegraphics[width=9cm,height=4.4cm,angle=0]{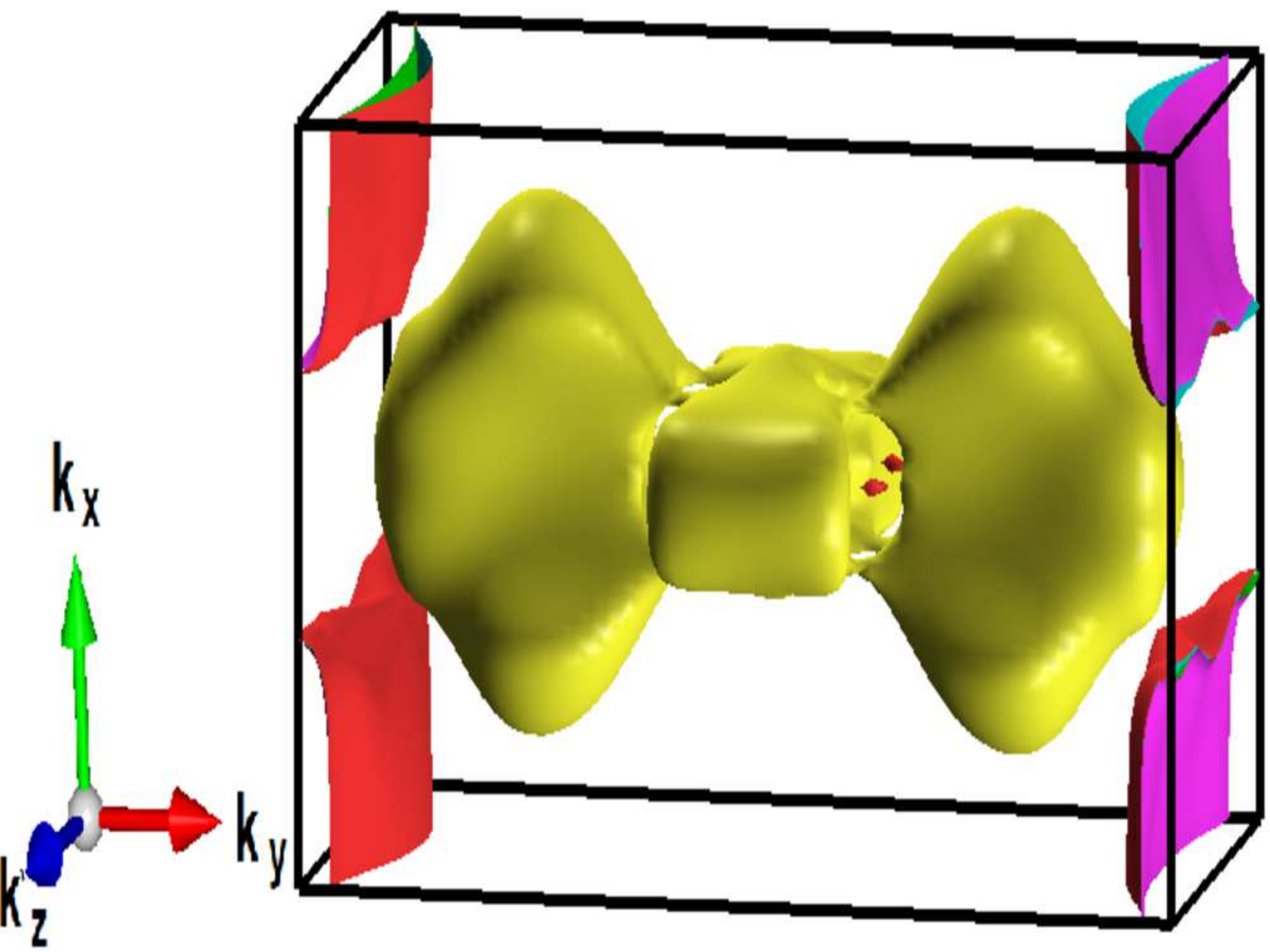}
\caption{Fermi surface of CrAs at the volume V2 (top panel) and V3 (bottom panel) in the first Brillouin zone.
}
\label{V3_FS}
\end{figure}

\begin{figure}[b!]
\centering
\includegraphics[width=9cm,height=4.4cm,angle=0]{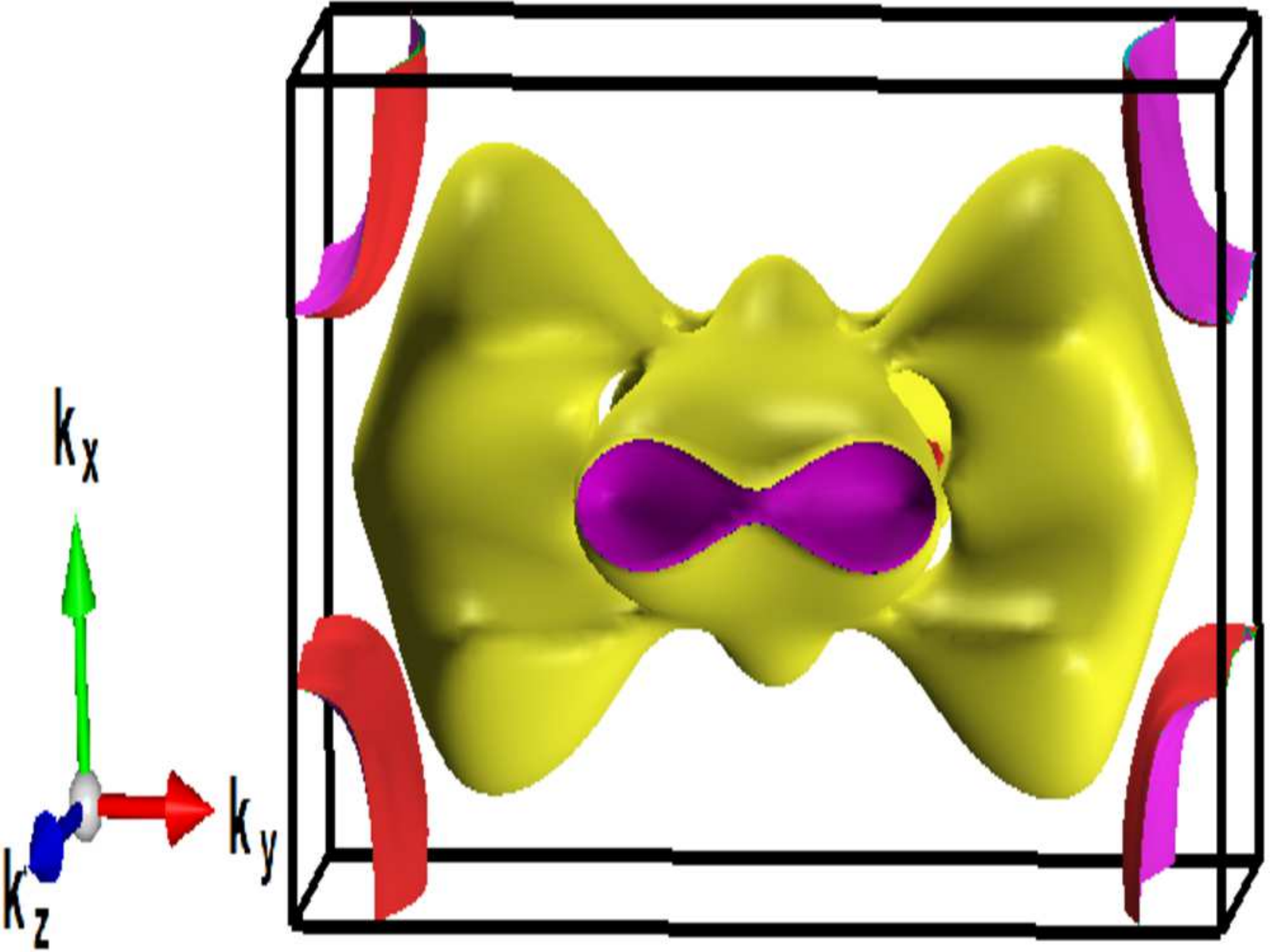}
\includegraphics[width=9cm,height=4.4cm,angle=0]{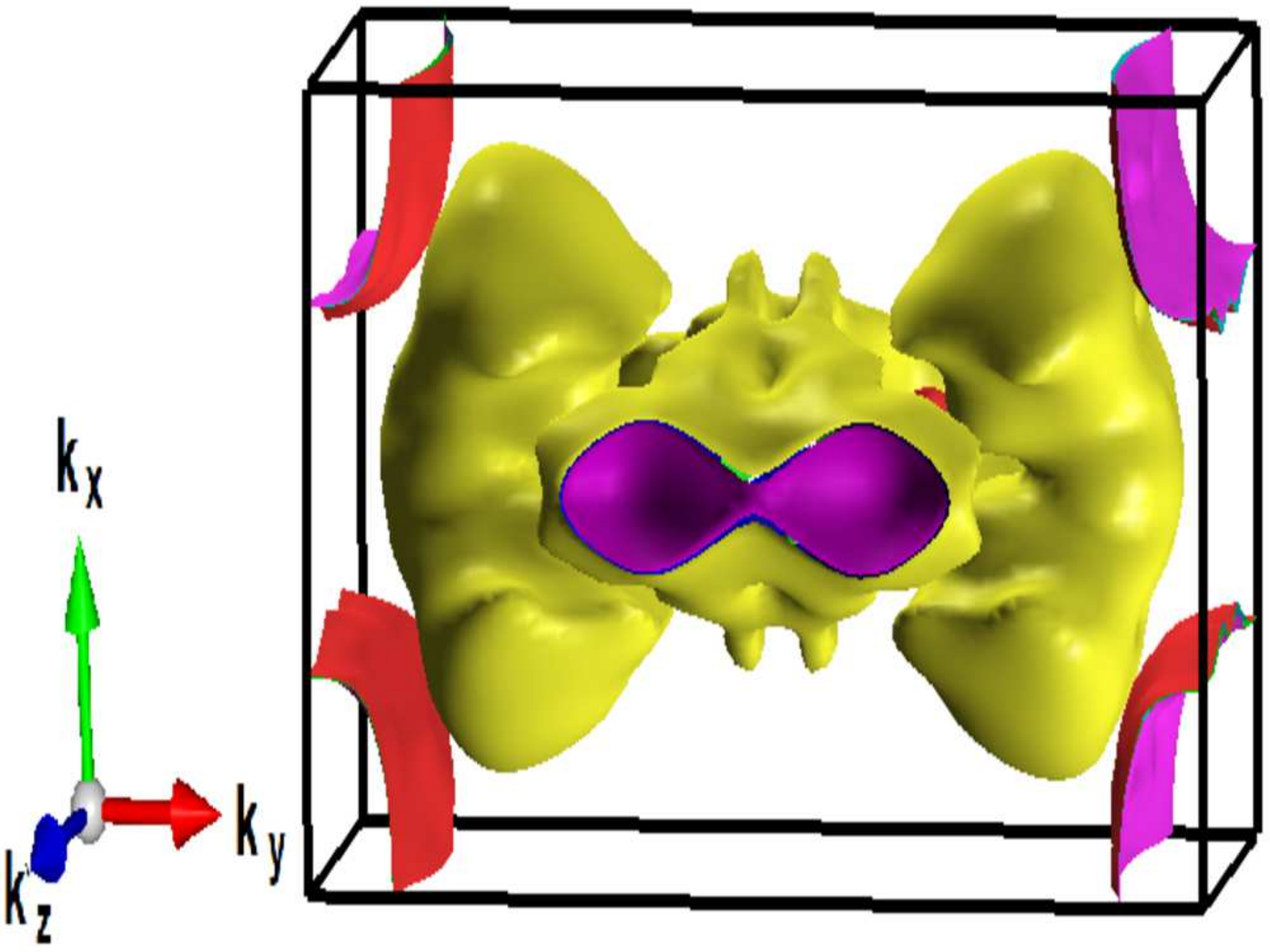}
\caption{Fermi surface of CrAs at the volume V4 (top panel) and V5 (bottom panel) in the first Brillouin zone.
}
\label{V5_FS}
\end{figure}

First of all, we may notice that from our calculation, as previously reported, it appears that the superconductivity is primarily associated with electrons of the Cr bands since these states dominate the DOS near the Fermi level.
Then, looking at Fig.~\ref{V1_FS} we see that the FS is composed by three cylinders; the red cylinder along the $k_x$-direction indicates the presence at the Fermi level of a two-dimensional band in the $bc$ plane. The other two concentric cylinders, at the border of the Brillouin zone, along the $k_z$-direction suggest the existence of other two-dimensional bands in the $ab$ plane. Moreover, we may also observe four small electron pockets.
When the volume of the unit cell gets reduced from V1 to V5, the shape of the FS strongly changes, as depicted in Figs.~\ref{V3_FS}-~\ref{V5_FS}. Indeed, we observe that the cylindrical shape of the FS found for the volume V1 disappears and the corresponding area of the FS increases.

\section{Discussion}

\begin{figure}[b!]
	\centering
	\includegraphics[width=8.6cm,height=6.3cm, angle=0]{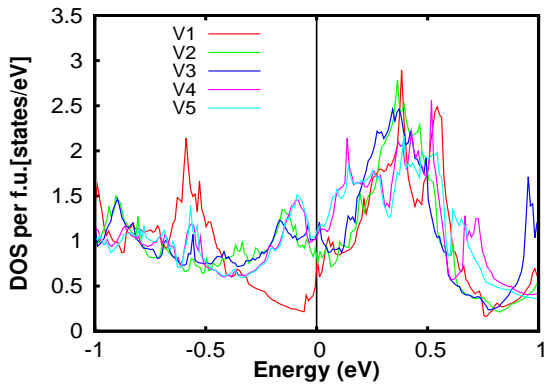}
	\caption{Total DOS near the Fermi level plotted for different volumes.}
	\label{DOSTOT}
\end{figure}

Let us comment on the connection between the results we presented in the previous Sections and experimental available data.
The DOS at the Fermi level increases with decreasing the volume, as shown in Fig.~\ref{DOSTOT} and summarized in Table~\ref{TabDOS}. We notice that the values of the DOS at the Fermi level for CrAs are lower than the ones found in FeSe and LaFeAsO compounds.~\cite{DSingh}
The trend we found possibly favors the superconducting transition as follows: we take into account the modification of the density of states $\rho(\varepsilon_F)$ at the Fermi level $\varepsilon_F$, considering that for a BCS-type superconductor the critical temperature and the order parameter are exponential functions of the amplitude of the DOS at $\varepsilon_F$  as $~exp(-1/\lambda \rho(\varepsilon_F) )$. Assuming that there is no change in the pairing coupling as well as in the pairing mechanism, the most significant effect on the strength of the superconducting order comes from the modification of the DOS at the Fermi level on the band where the electrons get paired. Hence, one would simply get an increase or a decrease of T$_c$ if the DOS at $\varepsilon_F$ grows or gets suppressed, as routinely applied in the analysis of the superconducting critical temperatures when superconductivity occurs in reduced dimensions and at the interfaces.~\cite{Dickey70,Strongin70,Li03, Bose05,Autieri12}
\begin{table}[!ht]
	\caption{We report the total density of state at the Fermi level $\rho(\varepsilon_F)$, as deduced from the calculated DOS plotted in Fig.~\ref{DOSTOT}, and the local spin moment $\mu$, as reported in Fig.~\ref{Moment}, for the volumes analyzed. The DOS is measured as states/eV and the local moment in unit of Bohr magneton.}
	\begin{center}
		\renewcommand{\arraystretch}{1.7}
		\begin{tabular}{|c|c|c|}
			\hline
			Volume      & $\rho(\varepsilon_F)$ & $\mu$     \\
			\hline
			V1              &    0.70   &  1.70 \\
			\hline
			V2              &    0.83   &  1.09 \\
			\hline
			V3              &    1.08   &  0.89 \\
			\hline
			V4              &    1.04   &  0.29 \\
			\hline
			V5              &    1.10   &  0.00 \\
			\hline
		\end{tabular}
	\end{center}
	\label{TabDOS}
\end{table}
\noindent Furthermore, the growth of the DOS at the Fermi level may also lead to a magnetic instability supporting the possible occurrence of a (antiferro)magnetic transition which could exhibit a non-trivial interplay with the superconducting phase.
It is worth noticing from Table~\ref{TabDOS} that the calculated density of states at the Fermi level for different volumes indicates that the critical superconducting temperature would increase with decreasing the volume of the unit cell from V1 to V5, simulating the effect of the external pressure. Moreover, the magnitude of the DOS follows an opposite trend with respect to the magnetic moment, suggesting a close relationship between these two quantities, further supporting our interpretation about the magnetic properties of CrAs.

Considering the behavior of the FS as a function of the volume of the unit cell, shown in the previous Section, the following picture may be inferred: when the volume of the unit cell gets reduced, the FS passes from a cylindrical shape to a closed one with a simultaneous  decrease of the magnetic moment.
Since the cylinder along the $k_x$-direction is due to the broken symmetry along the XS line in the $k$-space, the antiferromagnetic order and the spin magnetic moment are related to the modification of the FS.
Thus, the transition from two-dimensional FS to a three-dimensional one gives rise to a transition towards the non-magnetic phase of CrAs. We would like to notice that this modification takes place when the system goes into the superconducting phase, strongly suggesting that the superconductivity in this material is driven by the change of the FS from the two-dimensional shape to the three-dimensional one, accompanied also by the reduction of the magnetic moment.

\section{Final remarks}
In this paper we have investigated the structural, the magnetic and the electronic properties of CrAs using {\it ab initio} density functional theory. To simulate the role of the external pressure, we have varied the volume of the unit cell. The numerical simulation shows that when the volume increases, the system passes from a non-magnetic phase to the antiferromagnetic one, following the experimental available data.
We have found strong antiferromagnetic exchange at low pressure that weakens in the superconducting region.
Moreover, the obtained Fermi surface indicates that the system exhibits a two-dimensional behavior in $ab$ and $bc$ planes for larger values of the volume of the unit cell getting towards a three-dimensional shape when the volume of the cell decreases. This behavior, when considered in connection with the variation of the DOS at the Fermi level and of the magnetic moment, suggests a close link between the appearance of the superconductivity and the external pressure. Furthermore, the fair values for the magnetic moment found without the inclusion of the Coulomb interaction indicate that the CrAs may be considered as a weakly correlated material.

\noindent Concerning the properties of the superconducting state, it is well-known that almost all superconductors are conventional s-wave superconductors, mediated by electron-phonon interactions, and this may also be the case here. Indeed,  there is an experimental indication of conventional electron-phonon coupling in CrAs coming from the results related to the macroscopic phase separation of the magnetic and the superconducting phases.~\cite{khasanov15} The results suggest that the pressure-induced transition of CrAs from a magnetic to a superconducting state is characterized by a separation in macroscopic size magnetic and superconducting volumes, pointing towards an isotropic s-wave symmetry of the superconducting order parameter driven by electron-phonon interaction. This conclusion is further supported by the temperature behavior of the superfluid density that has found to scale with the critical temperature as $T_c^{3.2}$.
Moreover, since the Cr electrons dominate the DOS near the Fermi level, our results indicate that the superconductivity may verisimilarly be associated with the chromium sublattice.

To get more insight about the superconductivity mechanism, we propose to dope the CrAs and look at the critical temperature. In the framework of our results, from an inspection to the DOS plotted in Fig. \ref{DOSTOT}, and in a rigid band picture, we infer that a shift of the Fermi level  $E_F$ will increase the number of states at $E_F$, definitely enhancing the transition temperature between normal and superconducting state.
Finally, we point out that since the CrAs system is strongly anisotropic, an accurate study is also needed to take into account the related anisotropy in the phonon spectrum.~\cite{Aperis2015}

As a final consideration, we would like to point out that by using a method that combines the tight-binding approximation and the L\"{o}wdin down-folding procedure an effective Hamiltonian model describing the Cr d bands near the Fermi level has been derived.~\cite{autieri2017} To get this model Hamiltonian, as a first
step, a tight-binding model based on the Wannier transformation of the {\it ab initio} results here presented has been used. Hence, the energy spectra, the Fermi surface, the density of states and transport and magnetic properties of CrAs have been evaluated. The obtained results are shown to be consistent with the present {\it ab initio} calculations, as well as with the available experimental data for resistivity and Cr magnetic moment.

\section*{Acknowledgments}
We are grateful to J. Luo for stimulating discussions and A. Aperis, G. Cuono, F. Forte and A. Romano for useful discussions and critical reading of the manuscript.

\end{document}